\documentstyle[12pt]{article}

\begin{document}
\begin{center}

{ \bf
ABSTRACTS FOR THE JUNE 27-29, 1994 Princeton Conference}

{\bf
DISCRETE MODELS FOR FLUID MECHANICS}

%To receive anelectronic copy of this list, send the following
%e-mail message with a blank message body:
%mail comp-gas@xyz.lanl.gov
%Subject: get announce
\end{center}

\begin{center}
{ \bf
Particle Hydrodynamics
}\\
{
Berni Alder
\\}
{\it LLNL, P.O. Box 808, LLNL\\
Livermore, CA 94550, USA
}

\end{center}
{
\noindent
      Several particle methods such as molecular dynamics, lattice gas,
    lattice Boltzmann and Bird will be compared to Navier--Stokes
    continuum hydrodynamics as to their relative advantages and
    disadvantages. Particularly, the simulation of the ``consistent"
    Boltzmann and its Enskog extension will be discussed.
}

\begin{center}
{ \bf
     Bubble Dynamics Model for Hydrodynamically\\ Unstable Interfaces}\\
{
  U. Alon(1,2), J. Hecht(1), D. Ofer(1) , D. Mukamel(2) and D. Shvarts(1,3)
\\}

{\it

                    (1) Physics Department,
                  Nuclear Reasearch Center Negev\\
                 P.O.Box 9001, Beer-Sheva, Israel\\

                    (2) Physics Department,
                   Weizmann Institute of Science\\
                      Rehovot 76100,Israel\\

               (3) Laboratory for Laser Energetics,
                     University of Rochester\\
                       250 East River Road\\
                    Rochester, NY 14623, USA
}

\end{center}
{
\noindent
Evolution of the Rayleigh--Taylor unstable interface is known to be dominanted
by bubble growth and merging processes (1,2). Models that have been developed
describe the unstable front as composed of many bubbles (the basic particles)
which are rising in the gravitational  field (the one--body dynamics) and
merging (the two--body dynamics), where larger bubbles absorb smaller bubbles.
A new statistical model has been developed(3,4). The new model is based on the
Sharp--Wheeler--Glimm models(1), and incorporates the basic feature of
single--bubble rise and two--bubble competition using an extended Layzer
potential flow model for large density ratio($A=1$) and incompressible fluids.
The model offers a unified treatment of the Rayleigh--Taylor and the
Richtmyer--Meshkov instabilities. The merger model dynamics for both
instabilities are
shown to reach a scale invariant regime. For the RT instability, the model
predicts a constant acceleration of the front, growing as $0.05gt^2$, while for
the RM instability the model predicts a new novel scaling behavior for the
bubble front, growing as $at^{0.4}$. The model results are in good agreement
with experimental and 2d hydrodynamic simulations.\\

\noindent References:\\
\noindent 1. D.H.Sharp, Physica 12D, 3 (1984); C. L. Gardner, J. Glimm, O
McBryan,
   R. Menikoff, D. H. Sharp and Q. Zhang, Phys. Fluids 31, 447(1988).\\
2. J. A. Zufiria, Phys. Fluids 31, 440(1988).\\
3. U. Alon, J. Hecht, D. Mukamel and D. Shvarts, "Scale Invariant Mixing
   Rates of Hydrodynamically Unstable Interfaces"(submitted for publication).\\
4. U. Alon, D. Shvarts and D. Mukamel, Phys. Rev. E 48, 1008(1993).

}

\begin{center}
{ \bf Multiphasic Lattice Gas Models:\\ Prediction
of Surface Tension\\
and Growth Exponents}\\

{
C\'ecile Appert
\\}

{ \it Laboratoire de Physique Statistique\\ CNRS,
Ecole Normale Sup\'erieure,\\
24 rue Lhomond, 75231 Paris Cedex 05 France
}

\end{center}
{
\noindent
A liquid-gas phase separation is obtained for
a lattice gas model by addition of an attractive
force between particles.
As the interaction does not derive from a
potential, the equilibrium densities cannot be
derived from the classical Maxwell's construction.
A dynamical approach allows to predict the density
profile across a flat interface. The corresponding
equilibrium densities and the surface tension value
are in agreement with measurements on simulations
[Appert, d'Humi\`eres, and Zaleski, preprint 1994].

\noindent
Multiphasic lattice gas models may be applied
to study spinodal decomposition. We have used
the boolean model of immiscible fluids proposed
by D.H. Rothman and J.M. Keller, and extended to 3D by
J. Olson. Some growth exponents and power spectra scalings
have been computed for several color ratios.
}

\vfill\eject

\begin{center}
{ \bf
Hydrodynamic Limit of Lattice Boltzmann Equation
}

{
R. Benzi
\\}

{\it
   Dipartimento di Fisica,
   Universita Tor Vergata\\
   Viale E. Carnevale, I-00173, Roma, Italia
}

\end{center}

{

\noindent
In this talk I propose a new method to study the hydrodynamic
limit of the LBE. The method employs an empirical computation of
the inertial manifold for a simplified version of LBE. From
the numerical results, it is possible to define the rate of
convergence of LBE to the Navier Stokes equations.

}

\begin{center}
{ \bf
Correlations and Renormalization in \\Lattice Gases with Chemical
Reactions
}\\

{
Bruce Boghosian
\\}

{\it Thinking Machines Corporation\\
245 First Street, Cambridge, MA 02142, USA
}

\end{center}
{
\noindent
 The effects of correlations in a lattice gas for the Schl\"{o}gl
model chemical reaction are analyzed using diagrammatic summation
methods.  Corrections to the equilibrium densities and diffusivities are
calculated and compared to numerical experiment.  The impact of the
correlations on the problem of getting the lattice gas to exhibit the
correct stoichiometry is elucidated.

}

\begin{center}
{ \bf
{Lattice gases with Non--local Interactions}}\\

{
Jean Pierre Boon and  Olivier Tribel
\\}

{\it
        Physique Nonlin\'eaire et M\'ecanique Statistique\\
        Universit\'e Libre de Bruxelles, Campus Plaine CP 231\\
        1050--Bruxelles, Belgique
}

\end{center}
{
\noindent
Lattice Gas Automaton (LGA) models with long range interactions
are constructed by means of a power law distribution governing
velocity configuration changes at distant nodes. Such non--local
interactions mimic the effects of attractive/repulsive forces in
fluid systems. The equilibrium and transport properties are
considered for single-- and multi--species LGA's.

}

\vfill\eject
\begin{center}
{ \bf
Correlations in LGA's Violating Detailed Balance
}\\

{
H.J. Bussemaker and M. H. Ernst
\\}
{\it
 Instituut voor Theoretische Fysica\\
   Universiteit Utrecht\\
   Princetonplein 5, Postbus 80 006\\
   3508 TA Utrecht, The Netherlands
\\
}
\end{center}
\noindent  Non-detailed balance lattice gases exhibit strong velocity
correlation in equilibrium, as measured in FCHC and FHP
models. Here we present a kinetic equation, based on the
BBGKY hierarchy for LGA's, from which these equal time
velocity correlations can be calculated.
As a quantitative test we apply the theory to a fluid-type
LGA model on a triangular lattice and to a model
of interacting random walkers on a square lattice, both
violating detailed balance.
The numerical predictions agree very well with computer
simulations.

\begin{center}
{ \bf Exact Solutions for a Semi--Continuous\\ Model
of the Boltzmann Equation}\\

{
Henri Cabannes
\\}

{ \it Laboratoire de Mod\'elisation en M\'ecanique, associ\'e au CNRS\\
Universit\'e Pierre et Marie Curie\\
4 Place Jussieu, 75252 Paris Cedex 05 France
}

\end{center}
{

\noindent
The two--dimensional semi--continuous model of the
Boltzmann equation is an integro--differential equation, obtained
by assuming that all velocities have the same modulus ``c'',
but an arbitrary direction. For that model we built, as solitons, a family of
exact solutions. Those solitons are global solutions in time of the initial
value problem, and some of them correspond to partially negative initial data.
}

\def\beq{\begin{equation}}
\def\eeq{\end{equation}}
\def\beqar{\begin{eqnarray}}
\def\eeqar{\end{eqnarray}}
\def\nn{\nonumber}

\def\lsapprox{{\lower.3em\hbox{${\>\buildrel < \over \approx\>}$}}}

\begin{center}
{\bf Recovery of Complete Hydrodynamics in Lattice Gas Models}

{Hudong Chen }

{Exa Corporation}
\end{center}

\noindent We present theoretically that
the correct hydrodynamics including
its temperature evolution can be produced by a
lattice gas model in a systematic way.
The key steps are that a lattice gas system
must have an independent
energy conservation law and a proper dynamic
transition rate among different energy levels.
By a proper choice of the transition rate
and its functional dependence,
it is shown that the known non-Galilean invariant
artifacts in convection and pressure can be completely
removed in a lattice gas model.
In addition, we show that, with the removal of
the above artifacts, the model automatically gives rise to the
correct forms for the dissipative terms such
as the viscous stress tensor and the energy dissipation.
Therefore, the model produces the correct
hydrodynamic equations to all relevant orders.
Furthermore, it is shown, as a byproduct, the hydrodynamic temperature
and the thermodynamic temperature become equivalent.
The fundamental difference between this approach and
the approach used for recovering hydrodynamics
in the lattice Boltzmann method is that, in the former,
the proper equilibrium distribution is generated by a collision
process rather than being given a prescribed power series form.
Hence, it will stay positive-definite in any circumstance.

\begin{center}

{ {\bf An H--Theorem Without Semi--Detailed Balance}}

{Hudong Chen }

{Exa Corporation}
\end{center}

\noindent For a lattice gas system obeying Fermi-Dirac statistics but
not satisfying the so-called semi-detailed balance
condition, an H-theorem can still be proved if another
condition for the transition matrix is used.
Hence, the semi-detailed balance condition is
sufficient but not necessary condition for a system
to approach equilibrium.
It can be seen that the new condition has a broader
applicability to various lattice gas models including
those which have some phase transition properties.

\begin{center}
{\bf Growth Kinetics in  Multicomponent Fluids }

{Shiyi Chen and Turab Lookman}

{Theoretical Division T-13,

Los Alamos National Laboratory,

Los Alamos, NM 87545}

\end{center}

\noindent The hydrodynamic effects on the late stage
kinetics in spinodal decomposition of multicomponent fluids are examined using
a lattice Boltzmann scheme with stochastic fluctuations in the fluid and at the
interface. In two dimensions, the three- and four-component
immiscible fluid mixture
(with a $1024^2$ lattice) behaves like an off-critical binary
fluid with an estimated domain growth of  $t^{0.4\pm.03}$ rather than $t^{1/3}$
as
previously predicted, showing the significant influence of
hydrodynamics. In three dimensions (with a $256^3$ lattice), we
estimate the growth as $t^{0.96\pm.05}$ for both critical
and off-critical quenching, in agreement with phenomenological
theory.

\begin{center}

{\bf  Thermal Lattice Bhatnagar-Gross-Krook Model Without Nonlinear
       Deviations In Macro-dynamic Equations }

{ Y. Chen, H. Ohashi and M. Akiyama}

 {Department of Quantum Engineering and Systems Science,
         University of Tokyo }

{7-3-1 Hongo, Bunkyo-ku, Tokyo, 113, Japan}

\end{center}

\noindent
We present a new thermal lattice BGK model in $D$ dimensional space
for the numerical simulation of fluid dynamics. This model uses a
higher order velocity expansion of Maxwellian type equilibrium
distribution. In the mean time, the lattice symmetry has been
upgraded to ensure isotropy for the 6th order tensor of velocity
moments. These manipulations lead to macroscopic equations without the
nonlinear deviations, from which conventional thermal or non-thermal
lattice BGK models suffered. We demonstrate the improvements by
conducting classical Chapman-Enskog analysis and by the numerical
calculation of the structure of the shock wave front and the decaying
rate of the kinetic energy in the shear wave flow. Parameters in
the velocity expansion are given for example models in one, two and
three dimensions. The transport coefficients of the modeled $1D$ and
$2D$ fluids are numerically measured as well.

\begin{center}
{{\bf New Developments in Diffusion in }}\\
{{\bf Lorentz Lattice Gas Cellular Automata}}
\  \\
{E.G.D. Cohen} and { Fei Wang}\\
{\normalsize{\em Department of Physics, Rockefeller University,\\
1230 York Avenue, New York, NY 10021, USA}}
\end{center}

\noindent
Extending the calculations of the diffusion coefficient and related quantities
of a
Lorentz Lattice Gas to much longer times than before, a different diffusive
behavior was observed than reported before$^{[1,2]}$. One of the most striking
features is that the percolation transition does not seem to have observable
dynamical consequences. A survey of the present situation will be given.

---------------
\begin{enumerate}
\item X. P. Kong, E. G. D. Cohen, {\em Phys. Rev. B},
                {\bf 40}, 4838 (1989).
\item X. P. Kong, E. G. D. Cohen, {\em J. Stat. Phys.},
                {\bf 62}, 737 (1991).
\end{enumerate}

\centerline{\bf Cellular Automata Modeling of Reaction--Diffusion Phenomena}

\centerline{ Stephen Cornell, Michel Droz}

\centerline{ Department of Theoretical Physics, University of
Geneva, 24 quai Ernest-Ansermet.}
\baselineskip=10pt
\centerline{ CH-1211 Geneva 4, Switzerland}
\vglue 10pt
\centerline{\tenrm and}
\vglue 10pt
\centerline{ Bastien Chopard}
\baselineskip=12pt
\centerline{ Group for Parallel Computing,
University of Geneva, 24
rue General-Dufour.}
\baselineskip=10pt
\centerline{ CH-1211 Geneva 4, Switzerland}
\vglue 20pt

\noindent
Reaction-diffusion systems have long been a subject of interest
because of the complex behavior they exhibit, such as pattern
formation, chemical waves, etc.  The classical approach to this
type of problem is to use mean-field-like partial differential
equations, which are known not to describe the true behavior in
low dimensions.

\noindent
We present a method for simulating the microscopic dynamics of
reaction-diffusion systems using probabilistic cellular-automaton
algorithms.  A small number of random bits at each site produce
random walks, and simple synchronous update rules permit highly
efficient implementation using multi-spin coding techniques and/or
massively parallel machines with SIMD architecture.  Moreover, these
algorithms faithfully simulate the microscopic stochastic dynamics
(in the limit of low density, when lattice effects become unimportant).
We are therefore able to perform very precise investigations into the
departure from mean-field-like behavior of such systems below their
upper critical dimension.

\noindent
We study the effects of the initial condition on the evolution of
the system $m$A$+n$B$\to$ {\sl [inert]}, where $m$ and $n$ are small
integers.  In uniform  geometry, we give examples to show that,
for equal densities of A and B particles independently distributed,
the density decays like $t^{-x}$, with $x=\min(d/4, 1/[m+n-1])$,
whereas for correlated initial conditions the decay exponent is
$x=\min(d/2,1/[m+n-1])$.  For a time-independent reaction front
produced by opposing currents $J$ of $A$ and $B$ particles, we find
that the front is described by one length scale that behaves like
$J^{-\nu}$, with $\nu=\max(1/[d+1],[m+n-1]/[m+n+1])$.  If space is
divided into two regions, with constant initial densities of
A and B particles respectively, a time-dependent reaction front
evolves between the two regions.  This front exhibits scaling
behavior, with exponents related to those for the time-independent
case  described above, in contrast to the results reported by other
authors.

\vfill\eject

\def \l {\lambda }
\def \L {\Lambda }
\def \O {\Omega }
\def \p {\partial }
\def \o {\overline }
\def \v {\vert }
\def \g {\gamma }
\def \vphi {\varphi }
\def \t {\tilde }
\def \s {\sigma }
\def \d {\delta }
\def \th {\theta }
\def \z {\zeta }
\def \r {\rho }
\def \p0 {\parindent=0pt }
\def \p20 {\parindent=20pt }
\centerline {\bf Hexagonal Discrete Boltzmann Models}

\centerline {\bf With And Without Rest Particles }

\centerline { H. Cornille}

\centerline { Service de Physique th\'eorique, CE Saclay,
 F-91191 Gif-sur-Yvette, France}

\noindent
We compare the one hexagonal FHP$^{(1)}$ model without rest particle, with the
three
hexagonal GBL$^{(2)}$ model including rest particle and with a
five hexagonal model including also a rest particle. The models with and
without rest particles
were called ``thermal" and ``athermal" models by Ernst$^{(3)}$.
For shock waves, in the parameter space
building up the two equilibrium states (one parameter being the propagation
speed $\z $) satisfying the Rankine-Hugoniot relations, we study whether
different behaviors for the macroscopic quantities exist between
the two classes.
Firstly, from the knowledge of only the two equilibrium states associated
to ``shock profiles" solutions, we can predict$^{(4)}$ the subdomains where
the ratios $P/M$ ($P$ for pressure, $M$ for mass and $P/M$ for internal
energy) are monotonic or not.
Secondly, we investigate
the contour maps of the mass ratio $\rho $ across the shock and compare
with the corresponding one $\rho _C$ of the continuous theory$^{(5)}$ (valid
only for
models with an infinite number of velocities as was recently emphasized
by Cercignani$^{(6)}$).
Thirdly, we seek also the subdomains where
uniform solutions (densities associated to the same speed are the same
at the downstream state) can exist.
\parskip 10pt

\noindent
{\sl Our main result
is that for the thermal
models exist an additional subdomain including $\z \simeq 0$ with curve
$\rho =\rho _C$, possible nonmonotonic $P/M$ but without uniform solutions.}

\noindent
Recently for squares model$^{(4)}$, with and without rest particles, a similar
study
was performed. For both squares or hexagonal models the above results are in
agreement.
\parskip 10pt

\noindent {\bf References}

{\parskip 0pt

\noindent (1) U. Frisch, B. Hasslacher and Y. Pomeau,
Phys. Rev. Lett. 56, 1505 (1986).

\noindent (2) P. Grosfils, J. P. Boon and P. Lallemand, Phys. Rev. Let. 68,
107, 1992.

\noindent (3) M. Ernst, Adv. Math. Appl. Sc. 2,ed., A. S. Alves,
World Scientific 1991, p186, M. Ernst and
S. P. Das, J. Stat. Phys. 66, 465, 1992.

\noindent (4) H. Cornille, Saclay preprint 1993;
``VII Int. Conf. on Waves", ed. T. Ruggeri, Bologna, October 1993.

\noindent (5) H. Cornille, J. Stat. Phys. 48, 789, 1987.

\noindent (6) C. Cercignani, TTSP 23,1 ,1994.

}

\vfill\eject

\begin{center}
{ \bf Lattice Gas Models and Lattice Boltzmann\\ Equations for
Flows in Viscoelastic Media}

{Dominique d'Humi\`eres and Pierre Lallemand\\
Laboratoire de Physique Statistique \\
C.N.R.S., \'Ecole Normale Sup\'erieure \\
24, rue Lhomond, 75231 Paris cedex05, France\\

}
\end{center}

\noindent We shall present a family of lattice gas models displaying transverse
shear
waves$^{[1]}$. The corresponding lattice Boltzmann equation will be given.
The common feature of all these models is the conservation of additional
quantities related to the viscous stress tensor. It will be shown that
a suitable coupling between this new models and the ones with only mass
and momentum conservation leads to a viscoelastic behavior with transport
coefficients depending on the wave vector of typical perturbations.
The effects of ``spurious'' invariants and anisotropy will be considered.

\noindent [1] Dominique d'Humi\`eres and Pierre Lallemand, ``Gaz sur r\'eseau
pour
la repr\'esentation des ondes transverses'', {\it C. R. Acad. Sci. Paris II}
{\bf 317}, 997--1001 (1993).

\begin{center}
{ \bf Trends and Opportunities in Lattice Gas Research}\\

{
Gary Doolen
\\}

{\it
Theoretical Division, Los Alamos National Laboratory\\
Los Alamos, NM 87545, USA
}

\end{center}
{
\noindent Recent developments in applications of lattice Boltzmann
methods to problems of industrial interest will be discussed.
A probable Reynolds number upper limit of 10,000,000 for homogeneous
lattice, three-dimensional hydrodynamic simulations will be outlined.
Also, a summary of the usage of the lattice gas preprint library will be
given. Uses of bulletin boards, research frontier summaries, and multicasting
to accelerate lattice gas/lattice Boltzmann research will be presented.

}

\vfill\eject
\begin{center}
{ \bf
Chaos and Diffusion in Lorentz Lattice Gases
}\\
{
R.J. Dorfman
\\}

{\it
   Dept. of Physics, University of Maryland\\
   College Park, MD 20742, USA\\
}
{
M.H. Ernst
\\}

{\it
 Instituut voor Theoretische Fysica\\
   Universiteit Utrecht\\
   Princetonplein 5, Postbus 80 006\\
   3508 TA Utrecht, The Netherlands
\\
}
\end{center}
\noindent
We show that Lorentz lattice gases belong to the category of
dynamical systems with positive Lyapunov exponents, and are
therefore chaotic. As a result of the description of the Lorentz
lattice gas as a dynamical system, it is possible to relate a
macroscopic quantity characterizing irreversible behavior,
namely the diffusion coefficient, to microscopic dynamical
quantities, namely the Lyapunov exponents and the Kolmogorov-Sinai
entropy, of a set of trajectories that are trapped forever in
a finite region of the system. We show using arguments, based on
techniques from the kinetic theory of gases, that these dynamical
quantities can be explicitly computed and compared with the
results of computer simulations.

\begin{center}
{ \bf
Long-Range Correlations and \\Non--Gibbsian States in LGA's
}\\

{
M. H. Ernst and H.J. Bussemaker
\\}
{\it
 Instituut voor Theoretische Fysica\\
   Universiteit Utrecht\\
   Princetonplein 5, Postbus 80 006\\
   3508 TA Utrecht, The Netherlands
\\
}
\end{center}

\noindent
Non--Gibbsian stationary states occur in dissipative non--equilibrium
systems. They are closely connected with the lack of detailed balance,
and the absence of a fluctuation-dissipation theorem. These states
exhibit spatial and temporal correlations that are long-ranged under
generic conditions, even in systems with short range interactions,
provided the system has slow modes, and there is some anisotropy
either in configuration space or in state space.
The  anisotropy may come from imposed  fields
(driven diffusive systems, temperature gradients).
In the statistical mechanics of dissipative systems, such as
stochastic cellular automata, the asymmetry is only in state space.
Here the equilibrium states are non--Gibbsian, they may be  spatially
uniform with long range pair correlations, $g(r) ~ r^{-d}$ (with
dimensionality d), and may even exhibit instabilities, and lead to
the formation of clusters or patterns.

\vfill\eject
\begin{center}
{\bf Visualization of Dynamic Fluid Simulations:\\ Waves, Splashing, Vorticity,
Buoyancy}\\~\\\normalsize
Nick Foster and Dimitri Metaxas\\ Department of Computer and
Information Science,\\ University of Pennsylvania, Philadelphia, PA
19104\\
\end{center}

\noindent We develop a visualization technique for the
modeling and animation of viscous incompressible fluids. We use the
full time-dependent Navier-Stokes equations to comprehensively
simulate 2D and 3D incompressible fluid phenomena. Unlike previous
fluid animation graphics techniques which were based on approximating
solutions to fluid motion, we use these equations to accurately
model shallow or deep fluid flow, transient dynamic flow, vorticity
and splashing in simulated physical environments. In our simulations
we can also include variously shaped and spaced static or moving
obstacles that are fully submerged or penetrate the fluid surface.
We model interactions of moving obstacles with other moving or static
obstacles using our dynamic framework for modeling elastic and rigid objects.

\noindent
We use standard stable numerical analysis techniques based on
finite differences for the solution of the Navier-Stokes equations. We
solve these equations by iterating an initial set of pressures and
velocities defined over the finite-difference grid.  To model
free-surface fluids, we use a technique which is based on the
Marker-and-Cell method. Based on the fluid's pressure and velocities
obtained from the solution of the Navier-Stokes equations this
technique allows modeling of the fluid's free surface either by
solving a surface equation or by tracking the motion of marker
particles.  The later technique is suitable for visualization of
splashing and vorticity, since we render the marker particles as
spheres based on Silicon graphics hardware routines.

\noindent
Furthermore, we develop an editing tool for easy definition of a
physical-world which includes obstacles, boundaries and fluid
properties such as viscosity, initial velocity and pressure.  Using
our editor we can perform complex fluid simulations without prior
knowledge of the underlying fluid dynamic equations.  Finally,
depending on the application, we render the fluid's surface and the
other obstacles using well-developed graphics techniques to model
transparency, texture mapping, illumination,  reflectivity and
object material properties.  All of those techniques are implemented
using standard Silicon graphics hardware routines.

\noindent
We present a series of complex two- and three-dimensional simulations involving
fluid animations. The first two involve splashing phenomena caused by fluids
splashing into fluids, while the following three involve animations with
waves and interactions of fluids with obstacles including buoyancy.\\

\noindent 1) Fletcher, C.A.J., (1990) ``Computational Techniques for \\Fluid
Dynamics,''
Springer Verlag, Sydney.\\
\noindent 2) Metaxas, D., and Terzopoulos, D., (1992)
``Dynamic Deformation of Solid Primitives with Constraints,''
SIGGRAPH Proceedings, pp. 309--312.

\begin{center}
{ \bf Boundary Conditions in Discrete \\Kinetic Theory and Applications}\\

{
Ren\'ee Gatignol
\\}

{ \it Laboratoire de Mod\'elisation en M\'ecanique, associ\'e au CNRS\\
Universit\'e Pierre et Marie Curie\\
4 Place Jussieu, 75252 Paris Cedex 05 France
}

\end{center}
{

\noindent
The method of discretization of the velocities
allows to replace the usual Boltzmann equation by a system of
partial differential equations which is more tractable.
By using simple models without spurious invariants, we specify
the boundary conditions in two cases: first on an impermeable wall,
and second on an interface with the condensation and evaporation
phenomena. As applications, we consider Couette flows between two
parallel infinite plates, and the evaporation and condensation problems between
interfaces. With some particular models, we observe the phenomenon of
temperature inversion: the temperature is strictly increasing
from the hot interface to the cold interface.
}

\begin{center}
{{\bf
        First Steps towards a Description of Tracer Dispersion \\
        in Porous Media by Means of Lattice Gases
}}\\

{
        D. Grubert
\\}

{\it
NLfB-GGA (Geological Survey)\\
Stilleweg 2, D-30655 Hannover, Germany
}

\end{center}
{

\noindent
           The transport of particles in porous  media leads to
        dispersion, due to the tortuosity of the path. If the scale
        of the heterogeneity is much less than that of the experiment,
        it is possible to define an REV (Representative Elementary
        Volume) with a constant dispersivity. In this case, Fick's law
        is valid and the usual advection--diffusion equation can be
        applied.

\noindent   In strongly heterogeneous media, it is impossible to define
        a REV and the dispersion becomes scale-dependent. Therefore,
        as a different approach, flow and transport in a porous medium
        are modeled on a microscopic scale by means of lattice gases.

\noindent  First results of numerical experiments are presented, that
        simulate flow and dispersion for Hagen--Poiseuille flow and in
        a simple porous medium.

}

 \vfill\eject
\begin{center}
{ \bf
Internal Noise, Oscillations, Chaos\\ and Chemical Waves
}\\

{
Raymond Kapral
\\}

{\it
Chemical Physics Theory Group, Department of
Chemistry, \\ University of Toronto, Toronto, ON M5S 1A1, Canada
}

\end{center}

{\noindent
The effects of internal noise on chemical oscillations, chaos and
pattern formation processes will be described.  The problem
is investigated using a reactive lattice-gas automaton model
for the Willamowski-R\"{o}ssler reaction, a mass-action model that
exhibits a period doubling cascade leading to a strange attractor.
The nature of the stochastic dynamics is considered in both the
period-doubling regime, where the scaling structure is analyzed,
and in the regime where deterministic chaos is observed.
Pattern formation and wave propagation near deterministic
chaos are investigated in the fluctuating medium. The interplay
among spatial degrees of freedom, system size and internal fluctuations
are studied for this chaotic dynamical system.
}

\begin{center}

{\bf Cellular Automaton Model Of \\ Coupled Mass Transport
And Chemical Reactions }
{T. Karapiperis}
{\em Paul Scherrer Institute, CH-5232 Villigen PSI, Switzerland}
\end{center}
\vspace{-5mm}
\noindent
Mass transport, coupled with chemical reactions, is modeled as a cellular
automaton in which solute molecules perform a random walk on a regular lattice
and react according to a local probabilistic rule \cite{kb}. Stationary solid
particles dissolve with a certain probability and, provided solid is already
present and/or the solution is saturated, solute particles have a probability
to precipitate. No a priori restriction is placed on the number of particles
per lattice site. Assuming molecular chaos and a smooth spatial dependence of
particle density, we obtain, in the continuum limit, the macroscopic
reaction-transport equations with standard advection and diffusion terms.
The model is first applied to homogeneous systems
subject to the reactions $a + b \rightleftharpoons c$ and $a + b \rightarrow
c$,
respectively. In the reversible case we find that correlations between the
reacting particles can influence the macroscopic properties of the equilibrium
state (cf.\ \cite{db}). For the irreversible reaction, the long-time decline of
reactant density is slowed down because of density fluctuations that lead to
segregation of the reactants \cite{tw}. The rate of annihilation depends on
whether there is relative advection between the reactants. In the
simulation of an initially separated $a + b \rightarrow c$ system the width of
the reaction zone grows faster than predicted by the reaction-diffusion
equations \cite{gr}. The difference diminishes with enhanced diffusion, thus
suggesting a density fluctuation effect similar to the homogeneous case.
We also simulate autocatalytic reaction schemes
\cite{m} displaying spontaneous formation of spatial concentration patterns
(\cite{df}). Our simulation of the dissolution of a solid
block in a streaming fluid yields solid precipitation downstream from the
original solid edge, as a result of fluctuations in solute density.

\begin{center}
{ \bf
Simulation of Properties of Elastic Solids \\with Lattice Gases}\\

{\it
P. Lavall\'ee, D\'ept. Phy., Universit\'e du Qu\'ebec \`a Montr\'eal\\
 CP8888 Sta A, Montr\'eal H3C3P8, Canada\\
M. Kuntz, Geosciences Rennes, Campus de Beaulieu, 35043 Rennes C\'edex\\
J.C. Mareschal ,Geotop,Universit\'e du Qu\'ebec \`a Montr\'eal\\
CP8888 Sta A, Montr\'eal H3C3P8, Canada\\
}

\end{center}
{
\noindent
Although the equations for the deformation of a  linear elastic solid
are well known,  determination of the deformation and stress  is
difficult  except for very simple situations. The case where the elastic
solid is seeded with numerous microscopic defaults is particularly
difficult to deal with, even in two dimensions, both in the laboratory
and in numerical simulations. Experiments are very tedious since sample
preparation and interpretation of the results are time consuming, and
tests are usually destructive. Some simulations currently use analogies
with electrical resistance or springs networks.

\noindent Lattice gases provide an
alternative method of simulation. The basis of
the method is the equivalence principle, i.e.: the equations for the
displacement in a linear elastic solid are similar to the equations for
the velocity in an incompressible, Newtonian fluid undergoing creeping
motion. The patterns originating in a steady state fluid flow  with
appropriate boundary conditions can thus be interpreted to represent
patterns in an elastic solid and the macroscopic properties of solids can
 be measured.
The features of lattice gases can be exploited in this context because
lattice gases permit to simulate a large number of defaults.  A default
is conveniently simulated in a lattice gas as a region of very low
viscosity compared to that of the surrounding fluid:  a site identifying
a  default merely uses a different collision table than the surrounding
fluid. Solids containing up to 1000 (secant and non secant) defaults were
 simulated with simple shear conditions; the injection of particles was
arranged so that the total change of momentum required to maintain a
given velocity (displacement) gradient could be measured. The graph of
macroscopic properties of the solid (equivalent of Young modulus) as a
function of the number of defaults  was obtained and compared with
published results.

}

\begin{center}
{ {\bf Fluctuations and Chemical Waves\\ in Bistable Reacting System}}\\

A.T. Lawniczak

{\it Department of Mathematics and Statistics, University of Guelph

Guelph, Ont N1G 2W1, Canada
}

\end{center}
\noindent
A multi--species reactive lattice--gas automaton model is
used to study  the steady--state bifurcation structures, wave
propagation in a bistable chemical system, the Schl\"{o}gl model.
In contrast to eariler investigations of this system, the dynamics of
all three species that take part in the reaction are followed.
Far--from--equilibrium constraints are imposed by feeds of two
species from the boundaries, mimicking
 the situation that occurs in
continuously--fed--unstirred  reactors.
As a result, a detailed examination of the effects of
fluctuations and feed rate variations on
the steady--state structure and pattern formation processes can be
carried out.

\begin{center}
{ \bf
Shock Profiles in Lattice Models:\\ Some Exact and Some Simulation Results
}\\

{
Joel Lebowitz
\\}

{\it
Department of Mathematics and Physics\\
Hill Center, Rutgers University, New Brunswick\\
New Jersey, 08903, USA
}
\end{center}

{
\noindent
The full microscopic structure of macroscopic shocks is obtained exactly in
the uniform one-dimensional totally asymmetric simple exclusion process.
I shall also describe studies, via computer simulations, of the motion of shock
fronts in a
variety of other one- and two-dimensional stochastic lattice models with
parallel and serial dynamics, infinite and finite temperatures and
ferromagnetic and antiferromagnetic particle interactions.

\noindent The nonequilibrium stationary states of the asymmetric simple
exclusion
process with one site partially blocked were investigated both analytically
and via simulations.

}

\begin{center}

{\bf ALGE: A Supercomputer for 3-D Lattice-Gas Automata}

{ Fung Fung Lee and  Martin Morf}

      {   Computer Systems Laboratory}

       {  Stanford University}
        { Stanford, CA 94305}

         {lee@simd.stanford.edu}
\end{center}

\noindent
This paper presents the architecture of a scalable SIMD supercomputer,
called ALGE, designed for high Reynolds number 3-D lattice gas
simulations.  ALGE consists of an array of custom VLSI processors
interconnected as a 2-D torus.  Each processor has direct access only to
its local memory.  The effectiveness of ALGE is based on a set of
synergistic architectural mechanisms.  The most important one is the
collision unit.  By exploiting the group symmetry properties inherent in
the FCHC models, it reduces the size of the required collision lookup
table by more than 100 times without degrading Reynolds coefficients.
Other supporting architectural mechanisms include a novel address
generator containing a register file and multi-dimensional modulo adders
for efficient implementation of the virtual move update algorithm,
extensive random number generators for probabilistic boundary
conditions, transpose buffers for supporting a bit-serial word-parallel
organization and address sequences optimized for efficient DRAM
page-mode access, a simple and fast router for local communication,
support for statistics collection, and a scalable I/O mechanism.  The
speedup achieved by a large scale ALGE system is very close to linear.
A 2K-processor ALGE machine can be several hundred times faster than
current implementations on other supercomputers such as a 64K-processor
Connection Machine CM-2 or a 4-processor CRAY-2.
The future evolution possibilities of ALGE will also be discussed.

\vfill\eject
\begin{center}
{ \bf Turbulence In A Box}\\

{
Albert Libchaber
\\}

{\it
McDonnell Professor of Physics, Princeton University\\ Department of Physics\\
Jadwin Hall, P.O. Box 708, Princeton, NJ 08544--0708\\
and\\ Fellow, NEC Research Institute, 4 Independence Way, Princeton, NJ 08540\\
}

\end{center}
{
\noindent
Thermal turbulence in He gas at low temperature (5K) allows a broad study of
different
states of turbulence. By varying the gas density the Rayleigh number can be
swept
from $10^3$ to $10^{15}$ and the Reynolds number can go beyond $10^7$. All the
measured
quantities, in each turbulent state, scale with $R_a$ or $R_e$. We will
describe in
detail the characteristics of soft turbulence ($R_a < 10^8$), hard turbulence
($R_a > 10^8$), and present scaling models for various
models for various length scales and time scales in the problem.
Room temperature study in $SF_6$ gas allows a detailed analysis of the thermal
and velocity boundary layers. From such studies we will propose models for the
possible asymptotic state of thermal turbulence.
}

\begin{center}
{ \bf
Some Recent Developments in CFD\\ Based on Unstructured Grids}\\

{
Rainald L\"{o}hner\\
}

{\it
Institute for Computational Sciences and Informatics, MS 5C3\\
The George Mason University, Fairfax, VA 22030-4444\\
}

\end{center}
{
\noindent
The numerical solution of the (continuum) equations describing flows,
i.e. the Navier--Stokes equations, using unstructured, irregular
grids of triangles and tetrahedra, has seen a number of recent developments.
In particular, we will describe recent progress in the following areas:\\

\noindent - Generation of Navier--Stokes grids suitable for high-Re flows;\\
- Grid generation from discrete data;\\
- Edge--based flow solvers;\\
- Particle--Flow models and algorithms;\\
- Moving Grids;\\
- Parallelization;\\
- Link to CSD or CTD codes.\\

\noindent The talk will try to focus on the algorithmic aspects of each of
these areas, in the hope that a stimulating discussion with the
lattice gas CFD community can be established.

}

\begin{center}
{ \bf Physical Modeling on CAM--8}\\

{
N.H. Margolus
\\}

{\it Laboratory Computer Sciences\\
MIT, Cambridge, MA
}

\end{center}
{
\noindent
Conventional computers are grotesquely inefficient at running CA
models, and so discourage the development of these intrinsically
efficient kinds of computations.  To meet our own simulation needs,
and to encourage others to investigate CA models, we have developed
general--purpose CA machines which are optimized for studying
large-scale CA systems.  These machines are a small step towards
harnessing the kind of computational density and speed available in
nature only in a spatially local format.
Given a computing medium with a physics--like structure, it is only
natural to use it first to model physics itself.  In this talk, I will
demonstrate some of the physical modeling tasks that CA's have been
applied to on our new CAM--8 machines.  I will also discuss some of the
practical and theoretical modeling challenges that we are addressing
with the aid of our CA hardware.
}

\begin{center}
{ \bf
Comparison of Spectral Method and Lattice Boltzmann\\
 Simulations of Two-Dimensional Turbulence
}\\

{
William Matthaeus
\\}

{\it
Bartol Research Institute, University of Delaware\\
}
D. Martinez and S.Y. Chen\\
{\it
Theoretical Division, Los Alamos National Laboratory\\
}
D. Montgomery \\
{\it
Department of Physics and Astronomy, Dartmouth College\\
}

\end{center}
{
\noindent
Methods based on the Lattice Boltzmann Equation (LBE) approach are
an interesting and potentially useful alternative to traditional computational
methods for simulation of fluid turbulence.
The LBE method benefits from a considerable reduction
in noise relative to the Cellular Automaton (CA) fluid schemes upon which
they are based. In addition, recent advances in LBE methods alleviate or
completely
solve certain other problems that simple CA methods experience, while
maintaining
desirable features such as ease of implementation on vector and parallel
architectures.
In particular, LBE methods with pressure corrections and a single time
relaxation
approximation for collisions adhere to Galilean invariance, admit a well
behaved equation
of state, and rapidly approach local thermodynamic equilibrium in an easily
controllable
fashion. Using these recent advances, LBE fluid simulations schemes can be
meaningfully
compared with spectral method solutions to the equations of two dimensional
incompressible
hydrodynamics and magnetohydrodynamics (MHD). Comparisons of the methods for
hydrodynamics
show good (or, even excellent) quantitative agreement with regard to the
evolution of bulk
quantities such as energy and enstrophy, and also with regard to energy
spectra.
Detailed comparison of vorticity contours and streamlines indicate that very
similar
structures appear in the two simulation types. The main source of the computed
absolute
error (relative to the spectral simulation) appears to be a spatial drift of
these similar
structures. Examples of MHD LBE computation using a recently developed ``13
bit"
representation are also presented.
}

\begin{center}
{ \bf
       Thermal Lattice--Boltzmann Simulation\\ of Convective Flows
}\\

{
		         Guy McNamara
\\}

{\it
		    Center for Non--Linear Studies\\
		    Los Alamos National Laboratory\\
			    Los Alamos, NM\\

}

{
			     Alex Garcia
\\}

{\it
			Department of Physics\\
		      San Jose State University\\
			     San Jose, CA\\
}

\end{center}

\noindent We have developed a three--dimensional, 21--velocity, 4--speed,
thermal
lattice--Boltzmann model and have tested it on a number of simple
two--dimensional flows, including Rayleigh--B\'enard and slot convection.
This model employs a cubic lattice with particles moving at speeds $0,
1, \sqrt{3}$, and $2$ lattice spacings per time step.  Isothermal boundary
conditions are implemented using an extension of the so--called
reflection--principle boundary conditions for the Navier--Stokes
equations.  Measurements of speed and attenuation rate of traveling
sound waves show excellent agreement with theory.  Simulations of
Rayleigh--B\'enard and slot convection (with temperature ratios between
the hot and cold sides of the system as high as 2:1) are compared
against results from conventional finite--difference codes and show
good agreement.

\begin{center}
{ {\bf Catalytic Interface Erosion }}

Hsin-Fei Meng \\and\\ E. G. D. Cohen\\
The Rockefeller University, 1230 York Avenue, New York, NY 10021

\end{center}

{{\noindent
We study new interface erosion processes: the catalytic erosion.
We consider two cases:
the erosion of a completely occupied lattice by one single particle
starting from somewhere inside the lattice,
as well as the kinetic roughening of an initially flat surface,
where ballistic or diffusion-limited particles erode the surface but
remain intact themselves.
Many features resembling realistic interfaces, for example, islands and inlets
are generated.
The eroded regions can have continuously tunable fractal dimension; in
addition,
a rich variety of the dependence of the surface width on the system size
is observed.
}}

\begin{center}

 {\bf Digital Physics: \\
A New Technology for Fluid Simulation}

{Kim Molvig }

{Exa Corporation}
\end{center}

\noindent Exa Corporation is developing a commercial package for fluid dynamics
applications based on its {\it Digital Physics} Technology -- an extension
of the Lattice Gas concept familiar to those in attendance at this
conference. The development leverages the 1000x inherent computational
advantage of Digital Physics over the floating point approximations to
Navier-Stokes.  Digital Physics goes beyond previously known lattice gas
theory in several significant ways that are necessary to the practical
application of the method: 1) All discreteness artifacts are removed such
that the system behaves with very high statistical accuracy like a true
continuum fluid, including thermal and trans-sonic behavior; 2) Collision
efficiency has been improved to the point that the mean free path can be
reduced to a small fraction of the lattice spacing, corresponding to
$R^*\sim 40$; 3) Allowance is made for a variable resolution grid -- also
free of discretization artifacts; 4) Noise has been reduced substantially
over traditional lattice gas methods making it easy to make practical
measurements of transient quantities; 5) A Shot Noise Theorem is built into
the dynamics to insure that whatever fluctuations remain have no effect on
the mean dynamics; 6) A Slip Boundary Layer technology has been developed
to give low resolution access to very high Reynolds number flow regimes
including separation.  These are the theoretical features of the
technology. In addition Exa is developing the hardware and software support
that its vision of a Fluid CAD (FCAD) product requires.  Exa has designed
and built a custom ASIC chip and board that plugs into a Silicon Graphics
workstation to provide the capability 5,000,000 cell simulations at a
performance level of 10 Cray YMP pipes.  Future parallel implementations
will increase performance by factors of a hundred and more.  An integrated
software environment based on Parametric Technology Corporation's
Pro/Engineer, allows the user to import or create a complex solid shape in
CAD, have it compile down to the lattice and decompose automatically for
parallel execution, and visualize the data in an SGI Explorer environment.
The talk will focus on theoretical developments, simulation results and
computer architectural advantages, and give a brief overview of Exa's FCAD
concept.

{Partial support provided by the Department of Defense Advanced Research
Projects Agency
under contract DABT63-93-C-0010}

\begin{center}
{ \bf Steady Supersonic Flow in\\ a Discrete--Velocity Gas}\\

{
B.T. Nadiga
\\}

{\it MS--B258, CNLS\\
Los Alamos National Laboratory\\
Los Alamos, NM 87545, USA
}

\end{center}
{
\noindent
Unsteady shocks are easily generated in any discrete--velocity gas (DVG)
model and can be understood as serving to put the different parts of
the fluid in communication. However, the limited span of the
particle velocities in many DVG models preludes steady supersonic flow.
A DVG model with 25 velocities in 2D and a simple extension of it to 3D
is considered. The model has detailed balance and a study of the
characteristics
of the model--Euler equations shows that the model supports {\it steady}
supersonic
flow. A preliminary study of stationary shocks in the model is carried out
using a robust finite--volume flux--splitting scheme for the discrete--velocity
gases.
}

\begin{center}
{ \bf
The Global Solvability for Discrete Models}\\
\vspace*{24pt}
N.A.Nurlybaev\\
\vspace*{8pt}
{\it
Institute of Pure and Applied Mathematics\\
\vspace*{-3pt}
National Academy of Sciences\\
\vspace*{-3pt}
Almaty, Republic of Kazakhstan, 480021\\
}
\vspace*{6pt}
\end{center}

\noindent A simple and effective approach for obtain the a priori estimates
for discrete kinetic models is suggested\footnote
{N.Nurlybaev, Discrete velocity method in the theory of kinetic
equations, TTSP 22(1), 109-119(1993).}.
This approach is based on
one simple lemma which allows us to get positiveness of solution for
the Boltzmann equation and its discrete models. For some class of
discrete models in case of any dimensions and positive initial data
the boundedness of solution and accordingly the global existence
theorem are received.

\vfill\eject
\begin{center}
{ {\bf Fluid Mechanics and Statistical Mechanics:
\\Some Interesting Problems}}\\

{
Yves Pomeau
\\}

{\it Laboratoire de Physique Statiatique\\
CNRS, Ecole Normale Sup\'erieure,\\
24 rue Lhomond, 75231 Paris Cedex 05 France\\
and\\ Dept. of Mathematics, University of Arizona, Tucson USA.
}

\end{center}
{
\noindent
Among the many numerical methods in fluid mechanics, cellular automata and
Boltzmann lattice gases have the rather unique property of incorporating some
information related to molecular dynamics. Quite early in the development
of the subject this has been used to model combustion instabilities and
two--phase flows, both notoriously difficult to simulate with
conventional numerical methods. Other areas as well deserve to be
investigated from this point of view too. I shall consider as an example
of this situation the trans-sonic flows: as the previous examples,
the usual numerical methods meet problems there because the equations are
hyperbolic somewhere and elliptic somewhere else, making an uniform
``classical'' integration scheme very difficult to find (or very unstable...).
}

\vfill\eject
\begin{center}
{ \bf
Elimination of Staggered Invariants and Turbulence \\Simulations
with Lattice BGK Models
}\\

{
Yue--Hong Qian
\\}

{\it
211 Fluid Dynamics Research Center\\
Forrestal Campus, Princeton University\\
NJ 08544--0710, USA
}

\end{center}
{
\noindent
Spurious invariants exist in lattice gas and lattice Boltzmann models.
They may have important influence on the dynamics of the model
through the pressure term. A severe case
is the study of shock wave. We shall indicate possible ways of
eliminating staggered invariants for lattice BGK models.
As applications, we will show a simulation of shock wave and
present flow behaviors of high Reynolds numbers with or without
the elimination of staggered invariants.
}

\begin{center}
{ \bf
Lattice Boltzmann Model for Non--ideal \\Gases and their Mixtures}

{
Xiao Wen SHAN
\\}

{\it MS--B258, CNLS\\
Los Alamos National Laboratory\\
Los Alamos, NM87545, USA
}

\end{center}
{
\noindent
We will present in detail the previously proposed lattice Boltzmann
model for non--ideal gases and multiple components.  In this model,
interparticle forces are incorporated and the momentum is conserved
globally.  A non--ideal--gas equation of state can be derived in terms
of the interparticle interaction, and when properly chosen, a
liquid-gas type phase transition can be simulated.  The coexistence curve,
the density profile across the liquid--gas interface and surface tension
are all obtained analytically and are shown to be isotropic.  The
diffusivity in a mixture of non--ideal gases is also obtained through a
Chapman--Enskog calculation.
}

\vfill\eject
\begin{center}
{ \bf
    Lattice Boltzmann Computing: the Old Story and the
    New Perspectives
}\\

{
Sauro Succi
\\}

{\it
ECSEC--IBM, 171 Viale Oceano Pacifico\\
I--000141 Roma, Italy
}

\end{center}
{
\noindent
The basic ideas behind the Lattice Boltzmann
equation are presented together with an assessment of its
potential future developments.
}

% Typing aids
%
\newcommand{\vect}[1]{\mbox{$\bf #1$}}
\newcommand{\dbydt}[1]{\frac{{\rm D} #1}{{\rm D} t}}
\newcommand{\divg}{\mbox{\rm div}}
\newcommand{\Tr}{\mbox{${\rm Tr}$}}
\newcommand{\Or}[1]{\mbox{${\cal O}(#1)$}}
\newcommand{\eqn}[1]{(\ref{#1})}
\newcommand{\prim}[1]{{#1}^{\prime}}

\begin{center}
{ \bf Numerical Simulations
for Granular Flows Using Lattice BGK Models}

M-L. Tan, Y.H. Qian and S.A. Orszag

Fluid Dynamics Research Center\\
James Forrestal Campus\\
Princeton, New Jersey 08544-0710, USA
\end{center}

\noindent Many continuum theories for granular flow produce an equation of
motion for the fluctuating kinetic energy density (`granular
temperature') that accounts for the energy lost in inelastic
collisions.  Apart from the presence of an extra dissipative term,
this equation is very similar in form to the usual temperature
equation in hydrodynamics.  It is shown how a lattice-kinetic model
based on the Bhatnagar-Gross-Krook (BGK) approximation of the
Boltzmann equation that was previously derived for a miscible
two-component fluid may be extended to model the granular temperature
equation.  The extension is made by noting that in a two-species
mixture, the equation of motion for the concentration field for one
species can be made to be analogous to that for the granular
temperature when a source and a sink of concentration are added.  The
source and sink depend only on the local properties of the fluid and
correspond to the viscous heating and inelastic dissipation terms
respectively in the granular temperature equation.  A simulation of
the extended model for the case of an unforced granular fluid
reproduces the phenomenon of `clustering instability', viz. the
spontaneous creation of dense regions of low kinetic energy
interspersed in a dilute ambient of high energy, which occurs
generically in all granular flows.  The success of the continuum
theory in capturing this basic phenomenon, which lends credence to a
proposed physical mechanism for clustering in granular flows, is
discussed.

\vfill\eject
\begin{center}
{\bf Simulation Validations of Digital Physics}

{Chris Teixeira}

{Exa Corporation}
\end{center}

\noindent A summary of the theoretical attributes of Exa's Digital Physics
 simulation engine is presented that allows it to exactly emulate
 both the conservation and dissipative macroscopic behavior of
 single phase Newtonian fluid mechanics.  Simulation results for
 a series of benchmark fluid mechanics problems that demonstrate
 the quantitative
 validity of the model are presented including
 accurate reproduction of Coefficients of drag and Strouhal number
 for flow around circular cylinders for Reynolds number ranging
 from 10 to 1 million (including the drag crisis), accurate reproduction
 of coefficient of lift for a NACA airfoil for a range of increasing
 flow angles culminating with accurate capture of stall angle, and
 accurate reproduction of eddy-lengths beyond a backwards-facing step
 for a range of Reynolds numbers.  Simulation results will also
 be presented for a series of flows around geometries of practical
 interest and complexity including a realistic car body provided by
 Ford Motor Co. and a three-dimensional internal duct system provided
 by General Motors.

\begin{center}

{\bf An Engineering Fluid CAD Environment Based on Digital Physics }

{Kenneth R. Traub}

{ Exa Corporation}
\end{center}
\noindent
In this poster talk, we describe and demonstrate a complete, working software
environment for solving real-world fluid flow problems, based on Exa's {\em
Digital Physics\/} technology.  Digital Physics is a extension of lattice gas
technology, with these attributes of importance in solving complex flow
problems: removal of all discreteness artifacts, a variable resolution grid
technique, a greatly reduced level of statistical noise, and a boundary layer
simulation technique which gives low resolution access to very high Reynolds
number flows.  The unique attributes of Digital Physics are presented
elsewhere in this conference.

\noindent
The main components of Exa's Fluid CAD software system are: (1)~A front-end
for case construction built atop the Pro/ENGINEER mechanical CAD system from
Parametric Technology Corporation; (2)~An automated discretizer which converts
the description of a simulation case as a solid model into lattice form; (3)~A
high performance simulation engine which couples the Digital Physics
simulation algorithm with Exa's fluids accelerator hardware; and (4)~A
sophisticated fluid visualization environment built atop Explorer from Silicon
Graphics, Inc.

\noindent
We will demonstrate how a user can import a complex geometry into the CAD
front end, annotate it with flow properties including simple specifications of
variable resolution, submit the case for automatic discretization and
simulation, and explore the results in the visualization system.  We will show
real-world cases obtained from Exa's customers in the automotive industry.

\begin{center}

 {\bf Lattice Boltzmann Simulations of Laminar and\\ Turbulent Flow Past a
Cylindrical Obstacle}

 {\it Lukas Wagner and Fernand Hayot} \break
Department of Physics  \break
Ohio State University \break
Columbus, OH 43210 \break
\end{center}

\noindent We present lattice Boltzmann simulations of flow past a cylindrical
obstacle.  We are interested in the effect  of turbulence
in the incident flow on the coherence of the von Karman street.

\noindent Our study is based a L\'{e}vy walk model of turbulence in
a lattice Boltzmann model.
We discuss pressure around the cylinder in the laminar and
turbulent regimes, as well as the dependence of the von Karman
street on the analogue of integral scale in our model.

\begin{center}
{{\bf Diffusion in Honeycomb and Quasi}}\\
{{\bf Lorentz Lattice Gas}}  % (Title)
\  \\
{ Fei Wang} and { E.G.D. Cohen} \\
{\normalsize{\em Department of Physics, Rockefeller University,\\
1230 York Avenue, New York, NY 10021, USA}}
\end{center}

\noindent The diffusive behavior of a particle on a honeycomb and a
quasi-lattice
occupied randomly by fixed or flipping scatterers is discussed. A
comparison with the behavior on the square and the triangular lattices
will be made$^{[1,2]}$.

---------------
\begin{enumerate}
 \item E. G. D. Cohen, Fei Wang, The Fields Institute Series, {\em Am. Math.
Soc.} (1994).
 \item E. G. D. Cohen, in: Microscopic Simulations of Complex Hydrodynamic
Phenomena,
	NATO Series, {\bf 92}, M. Mareschal and B. L. Holian, eds., Plenum Press, New
	York, 137-152 (1992).
\end{enumerate}

\centerline{\bf  Hydrodynamical Simulations from BGK Model}

\centerline{ Kun Xu, Luigi Martinelli and Antony Jameson }
\centerline{Department of Mechanical and Aerospace Engineering}
\centerline{Princeton University }

\noindent Starting from the gas-kinetic Bhatnagar--Gross--Krook model
(BGK), a new gas--kinetic scheme for the
hydrodynamical compressible Euler and Navier--Stokes equations was
developed.

\noindent This scheme differs from any other ``Boltzmann--type schemes,''
in which the integral solution of the BGK model and the collisional
 conservation constraints are used at the same time. Also,
the continuous particle distribution function is adopted in the
calculation of time--dependent numerical fluxes.
In this paper, we give the basic idea of constructing this scheme and
analyze its properties in terms of central--difference and upwind
schemes. The similarities and differences with the
Lattice Boltzmann methods are mentioned.
At the end, some test cases for the 1--D and
2--D steady and unsteady gas flows are presented.

\begin{center}
{ \bf
Intermittency of Dissipation Rate in Turbulence
}\\

{
Victor Yakhot
\\}

{\it
214 Fluid Dynamics Research Center\\
Forrestal Campus, Princeton University\\
NJ 08544--0710, USA
}

\end{center}

\def\ZZ{Z\!\!\!Z\,}
\centerline{\bf Incompressible Limit of Discrete Velocity Model}
\centerline{  Horng-Tzer Yau }
\centerline{Courant Institute, New York University}
\centerline{New York, NY 10012}
\centerline{e-mail: yau@math.nyu.edu}

\noindent Consider the
discrete velocity model with random collisions  on $\ZZ^d$. The configurations
 at each site $x \in \ZZ^d$ consist of particles with unit velocity in each
direction.
There is a hard core
interaction among particles with the same velocity so that at each site
there is at most one particle with a given velocity. For example, in
$d=3$ there are six types of particles corresponding to the six unit vectors
in $\ZZ^3$. The dynamics consists of two parts: the asymmetric simple exclusion
part
and the collision part. For particle at $x \in \ZZ^d$  with velocity $e$,
 it jumps to position $x+e^{\prime}$ with rate $ \alpha_e (e^{\prime})$
 provided the jump does not lead to a violation with the hard core interactions
among particles with the same velocity. In other words, particle at $x$ with
velocity $e$
will jump to $x+e^{\prime}$ only if there is no particle at $x+e^{\prime}$ with
velocity
$e$. The velocities of the particles are preserved
 under this dynamics so that particles with velocity $e$ remain  particles
 with velocity $e$ after the
jump. If $ \alpha_e (e^{\prime}) = \delta(e ,e^{\prime})$, particles with
velocity $e$ jump with
rate 1 in the direction $e$ and the identification of these particles as having
velocity $e$ is appropriate. Otherwise it should be considered as having
velocity
$\alpha_e (e^{\prime})$. This is the asymmetric simple exclusion part of the
dynamics.

The collision part can be described as follows. If particles with
velocities $e$ and $-e$ occur simultaneously at any site $x \in \ZZ^d$, the
velocities of particles
change to $e^{\prime}$ and $-e^{\prime}$ with rate one for any direction
$e^{\prime} \not= e$,
again subject to no violation with the hard core interaction. This is the
collision part of the dynamics.

We consider the incompressible limit of this model in dimension $d \ge 3$,
namely, the average
density and velocity of particles are of order $\epsilon$ while the time scale
of the dynamics
is of order $\epsilon^{-2}$. We prove that the empirical velocity (i.e., the
macroscopic
velocity of the particle systems) satisfies a Navier-Stokes equation. The
viscosity of the
resulting equation is
characterized by a Green-Kubo formula, which is then reformulated as a
variational principle.

\end{document}